\documentclass[12pt,twoside]{article}
\usepackage[mathscr]{eucal}
\usepackage{amsmath,amsfonts,amssymb,amsthm,mathabx,empheq}

\usepackage{times}
\usepackage{pdfsync}
\usepackage{cite}
\usepackage{url}
\usepackage{hyperref}
\usepackage{tensor}
\usepackage{color}

\allowdisplaybreaks[1]

\advance\textheight\topskip
\def\nn{\nonumber}

\def\p{\partial}

\def\n{\nabla}

\def\cG{\mathcal{G}}

\def\cO{\mathcal{O}}

\newcommand{\bea}{\begin{eqnarray}}
\newcommand{\eea}{\end{eqnarray}}
\newcommand{\be}{\begin{equation}}
\newcommand{\ee}{\end{equation}}

\numberwithin{equation}{section} \makeatletter
\@addtoreset{equation}{section}

\DeclareFontFamily{OT1}{rsfs}{} \DeclareFontShape{OT1}{rsfs}{m}{n}{
<-7> rsfs5 <7-10> rsfs7 <10-> rsfs10}{}
\DeclareMathAlphabet{\mycal}{OT1}{rsfs}{m}{n}

\newcommand*\xbar[1]{%
  \hbox{%
    \vbox{%
      \hrule height 0.5pt 
      \kern0.3ex
      \hbox{%
        \kern-0.0em
        \ensuremath{#1}%
        \kern-0.0em
      }%
    }%
  }%
}

\begin{document}

\title{Asymptotic structure of Einstein-Gauss-Bonnet theory in lower dimensions}

\author{ H. L\"{u} and Pujian Mao }

\date{}

\def\mytitle{Asymptotic structure of Einstein-Gauss-Bonnet theory in lower dimensions}

\pagestyle{myheadings} \markboth{\textsc{\small   H.~L\"{u}, P.~Mao   }}
{\textsc{\small Einstein-Gauss-Bonnet in lower dimensions}}

\addtolength{\headsep}{4pt}

\begin{centering}

  \vspace{1cm}

  \textbf{\Large{\mytitle}}

  \vspace{1.5cm}

  {\large     H. L\"{u}$^{\clubsuit}$ and Pujian Mao$^{\diamondsuit}$    }

\vspace{.5cm}

\vspace{.5cm}
\begin{minipage}{.9\textwidth}\small \it  \begin{center}
     Center for Joint Quantum Studies and Department of Physics,\\
     School of Science, Tianjin University, 135 Yaguan Road, Tianjin 300350, China
 \end{center}
\end{minipage}

\end{centering}

\begin{center}
$^{\clubsuit}$mrhonglu@gmail.com,\ \ \  $^{\diamondsuit}$pjmao@tju.edu.cn
\end{center}

\vspace{1cm}

\begin{center}
\begin{minipage}{.9\textwidth}

\textsc{Abstract}. Recently, an action principle for the $D\rightarrow4$ limit of the Einstein-Gauss-Bonnet gravity has been proposed. It is a special scalar-tensor theory that belongs to the family of Horndeski gravity. It also has a well defined $D\rightarrow3$ and $D\rightarrow2$ limit. In this work, we examine this theory in three and four dimensions in Bondi-Sachs framework. In both three and four dimensions, we find that there is no news function associated to the scalar field, which means that there is no scalar propagating degree of freedom in the theory. In four dimensions, the mass-loss formula is not affected by the Gauss-Bonnet term. This is consistent with the fact that there is no scalar radiation. However, the effects of the Gauss-Bonnet term are quite significant in the sense that they arise just one order after the integration constants and also arise
  in the quadrupole of the gravitational source.
 \end{minipage}
\end{center}


\thispagestyle{empty}
\newpage
\tableofcontents


\section{Introduction}
\label{sec:introduction}

Einstein-Gauss-Bonnet (EGB) gravity is the simplest case of Lovelock's extension of Einstein gravity \cite{Lovelock:1971yv}. The theory exists naturally in higher dimensions and becomes important with the development in string theory. Its black hole solutions \cite{EGBBH,cai} play an important role in studying AdS/CFT correspondence. In four dimensions, the Gauss-Bonnet combination is a topological invariant and does not affect the classical equations of motion. Einstein's General Relativity is widely believed to be the unique Lagrangian theory yielding second order equations of motion for the metric in four dimensions. The Lovelock type of construction requires additional scalar or vector fields, giving rise to Hordenski gravities \cite{Horndeski:1974wa} or generalized Galilean gravities \cite{galilean}.

However this has been recently challenged by a novel four dimensional Einstein-Gauss-Bonnet solution \cite{Glavan:2019inb}. The fresh treatment is encoded in the dimensional regularization. After a rescaling of the coupling constant $\alpha\rightarrow\frac{\alpha}{D-4}$, the $D\rightarrow4$ limit can be taken smoothly at the solution level, yielding a nontrivial new black hole. This created a great deal of interest \cite{alot}, as well as controversy \cite{Gurses:2020ofy}, as one would expect that higher-derivative theories of finite order that are ghost free in four-dimensions cannot be pure metric theories but are of the Hordenski type. In fact the resolution of the divergence at the action level is far less clear and the action principle for the $D=4$ solution is not given in \cite{Glavan:2019inb}. One consistent approach is to consider a compactification of $D$-dimensional EGB gravity on a maximally symmetric space of $(D-p)$ dimensions where $p\leq4$, keeping only the breathing mode characterizing the size of the internal space such that the theory is minimum. The $D\rightarrow p$ limit can then be smoothly applied \cite{Lu:2020iav}, leading to an action principle admitting the four dimensional Einstein-Gauss-Bonnet solution \cite{Glavan:2019inb,Cai:2009ua} (see also \cite{Kobayashi:2020wqy,Hennigar:2020lsl}). In fact the analogous $D\rightarrow 2$ limit of Einstein gravity was proposed many years ago \cite{2d1} (see also the recent work \cite{2d2}). It turns out that the resulting theory is indeed a special Horndeski theory. The action contains a Horndeski scalar that coupled to the Gauss-Bonnet term, as well as the metric field. The lower dimensional action is given by \cite{Lu:2020iav} \footnote{We have chosen that the curvature tensor of the internal maximally symmetric $D-p$ space vanishes.}
\be\label{lagrangian}
S_p=\int d^p x \sqrt{-g} \bigg[R + \alpha \phi \cG + \alpha \big( 4 G^{\mu\nu} \n_\mu \phi \n_\nu \phi - 4 (\n \phi)^2  \n^2 \phi + 2(\n \phi)^2(\n \phi)^2 \big) \bigg].
\ee
where $G^{\mu\nu}$ is the Einstein tensor, and
\be
\cG\equiv R^{\mu\nu\rho\sigma}R_{\mu\nu\rho\sigma} - 4 R^{\mu\nu} R_{\mu\nu } + R^2
\ee
is the Gauss-Bonnet term.

There are several interesting features in the new theory \eqref{lagrangian}. First, there is no scalar kinematic term. So scalar propagator should be absent. Second, the classical solution of the Minkowski vacuum admits two independent scalar solutions namely $\phi=0$ which we refer to as the ordinary vacuum and $\phi=\log \frac{r}{r_0}$ which we refer to as the logarithm vacuum.\footnote{We set a constant radial scale $r_0$ to compensate the length dimension in $\phi$.} Last but not least, the $\alpha$ correction is inherited from their higher-dimensional counterparts. Hence it includes not only the four dimensional Gauss-Bonnet term coupled with a scalar field but also scalar terms that are non-minimally coupled to gravity. The latter seems to be more significant than the former in the corrections to the classical solution of Einstein gravity.

To test the above interesting features, we will study the asymptotic structure of the lower dimensional EGB theory \eqref{lagrangian} in the Bondi-Sachs framework \cite{Bondi:1962px,Sachs:1962wk} in the present work. In 1960s, Bondi et al. established an elegant framework of asymptotic expansions to understand the gravitational
radiation in axisymmetric isolated system in Einstein theory \cite{Bondi:1962px}. The metric fields are expanded in inverse powers of a radius coordinate in a suitable coordinates system and the equations of motion are solved order by order with respect to proper boundary conditions. In this framework \cite{Bondi:1962px}, the
radiation is characterized by a single function from the expansions of the metric fields which is called the news function. Meanwhile, the mass of the system always decreases whenever there is news function. Shortly, Sachs extended this framework to asymptotically flat spacetime \cite{Sachs:1962wk}. It is a good starting point to study the asymptotic structure of the theory \eqref{lagrangian} in three dimensions. We obtain the asymptotic form of the solution space. There is no news function in three dimensions. This is a direct demonstration that there is no scalar propagating degree of freedom. Then we turn to the four dimensional case. Two scalar solutions of the vacuum lead to two different boundary conditions for the scalar fields. The solution spaces are obtained in series expansion respect to different boundary conditions. For both cases, there is no news function in the expansion of the scalar field which means that scalar propagating degree of freedom does not exist in four dimensions either. And the $\alpha$ corrections are transparent in the solution space. They arise just one order after the integration constants and also arise in the quadrupole, i.e. the first radiating source in the multipole expansion. In the logarithm vacuum, the $\alpha$ corrections even live at the linearized level. We show the precise formula of the $\alpha$ corrections in the quadrupole. Hence the two different vacua are indeed experimentally distinguishable.

The organization of this paper is quite simple. In the next section, we study the asymptotic structure in three dimensions. We perform the same analysis in four dimensions in section \ref{sec:4d} with special emphasis on $\alpha$ corrections in the gravitational solutions and the classical radiating source. After a brief conclusion and discussions on some of the future directions, we complete the article with one appendix where some useful relations are listed.

\section{Asymptotic structure of Einstein-Gauss-Bonnet theory in three dimensions}
\label{sec:3d}

As a toy model, it is worthwhile to examine the EGB theory \eqref{lagrangian} in three dimensions to see if the Bondi-Sachs framework is applicable to this theory. In three dimensions, the Gauss-Bonnet term is identically zero.  Applying the relations in Appendix \ref{appendix}, the variation of the action is obtained as
\begin{multline}
\delta S_3=\int d^3 x  \sqrt{-g}\bigg\{-\frac12 g_{\tau\gamma}\delta g^{\tau\gamma} \bigg[R + \alpha \big( 4 G^{\mu\nu} \n_\mu \phi \n_\nu \phi
 - 4 (\n \phi)^2  \n^2 \phi\\
  + 2(\n \phi)^2(\n \phi)^2 \big) \bigg] + R_{\mu\nu} \delta g^{\mu\nu} + \n_\mu \left(g_{\alpha\beta} \n^\mu \delta g^{\alpha\beta} - \n_\nu \delta g^{\mu\nu}\right)\\
 + \alpha \bigg[2 \left(g_{\rho\sigma} \n^\mu \n^\nu \delta g^{\rho\sigma} - \n_\rho \n^\mu \delta g^{\rho\nu} - \n_\rho \n^\nu \delta g^{\rho\mu} + \n^2 \delta g^{\mu\nu} \right) \n_\mu \phi \n_\nu \phi\\
 + 4 R^\mu_\rho \n_\mu\phi \n_\nu\phi \delta g^{\nu\rho} + 4 R^\nu_\rho \n_\mu\phi \n_\nu\phi \delta g^{\mu\rho} - 2 \bigg( R \n_\mu \phi \n_\nu \phi \delta g^{\mu\nu} + (\n \phi)^2 R_{\rho\sigma} \delta g^{\rho\sigma}\\
  +  g_{\rho\sigma} (\n \phi)^2 \n^2 \delta g^{\rho\sigma} - (\n \phi)^2 \n_\rho \n_\sigma \delta g^{\rho\sigma} \bigg)- 4 \delta g^{\mu\nu} \n_\mu \phi \n_\nu \phi \n^2 \phi \\
- 4 (\n \phi)^2 \n_\rho \n_\sigma \phi \delta g^{\rho\sigma}   + 4 \delta g^{\mu\nu} \n_\mu \phi \n_\nu \phi (\n \phi)^2
 + 2 (\n \phi)^2  g_{\mu\nu} \n_\rho \phi \n^\rho \delta g^{\mu\nu}\\
   - 4 (\n \phi)^2 \n_\rho \phi \n_\mu \delta g^{\rho\mu}   \bigg]
 + \alpha \bigg[ 8  G^{\mu\nu} \n_\mu \delta \phi \n_\nu \phi  - 8 g^{\mu\nu} \n_\mu \delta \phi \n_\nu \phi \n^2 \phi \\
 + 8 g^{\mu\nu} \n_\mu \delta \phi \n_\nu \phi (\n \phi)^2  - 4 (\n \phi)^2 \n^2 \delta \phi \bigg]\bigg\}.
\end{multline}
After dropping many boundary terms, one obtains the Einstein equation
\be
G_{\mu\nu} - \alpha T_{\mu\nu}=0,
\ee
where
\begin{multline}\label{t}
T_{\mu\nu}= g_{\mu\nu} \big[ 4 R^{\rho\sigma} \n_\rho \phi \n_\sigma \phi + 2 \n_\sigma \n_\rho \phi \n^\rho \n^\sigma \phi\\
  - 2 (\n^2\phi)^2 +  (\n \phi)^2(\n \phi)^2 + 4 \n_\rho\n_\sigma\phi \n^\rho\phi \n^\sigma \phi \big]\\
+ 4 \n_\mu \n_\nu \phi \n^2 \phi - 4 \n_\rho\n_\mu \phi \n^\rho \n_\nu \phi + 4 \n_\mu \phi  \n_\nu \phi \n^2 \phi - 4 \n_\rho \n_\mu \phi \n_\nu \phi \n^\rho \phi\\
 - 4 \n_\rho \n_\nu \phi \n_\mu \phi \n^\rho \phi - 4 \n_\mu \phi \n_\nu \phi (\n\phi)^2 - 4 R^\rho_\nu \n_\mu \phi \n_\rho \phi - 4 R^\rho_\mu \n_\nu \phi \n_\rho \phi\\
+2 R \n_\mu \phi \n_\nu \phi + 2 G_{\mu\nu} (\n \phi)^2 - 4 R_{\mu\rho\nu\sigma}\n^\rho \phi \n^\sigma \phi ,
\end{multline}
and the scalar equation
\begin{multline}
G^{\mu\nu} \n_\mu \n_\nu \phi + R^{\mu\nu} \n_\mu \phi \n_\nu \phi  + \n^2\phi (\n \phi)^2  - (\n^2\phi)^2\\
+ 2 \n_\rho \n_\sigma \phi \n^\sigma\phi \n^\rho \phi  +  \n_\rho \n_\sigma \phi \n^\sigma \n^\rho \phi = 0.
\end{multline}

\subsection{Bondi gauge}
In order to study three dimensional Einstein at future null infinity, Bondi gauge was adapted into three dimensions with the gauge fixing ansatz \cite{Barnich:2006av,Barnich:2010eb}
\be
\label{metric3d}
ds^2=\frac{V}{r} e^{2\beta} du^2 - 2 e^{2\beta} dudr + r^2 (d\phi - U du)^2,
\ee
in $(u,r,\varphi)$ coordinates and $\beta,U,V$ are functions of $(u,r,\varphi)$. 
Suitable fall-off conditions that preserving asymptotic flatness are
\begin{equation}
\label{boundary3d}
U=\cO(r^{-2}),\quad V=\cO(r), \quad \beta=\cO(r^{-1}),\quad \phi=\cO(r^{-1}).
\end{equation}

One of the advantages of the Bondi gauge is encoded in the organization of the equations of motion \cite{Bondi:1962px,Sachs:1962wk,Barnich:2010eb} (see also \cite{Barnich:2015jua,Lu:2019jus} for the generalization to matter coupled theories). There are four types of equations of motion, namely the main equation, the standard equation, the supplementary equation and the trivial equation. The terminologies characterize their special properties. The main equations determine the r-dependence of the unknown functions $\beta,U,V$ while the standard equation controls the time evolution of the scalar field. Because of the Bianchi identities, the supplementary equations are left with only one order in the $\frac1r$ expansion undetermined and the trivial equation is fulfilled automatically when the main equations and the standard equation are satisfied. In three dimensional EGB theory \eqref{lagrangian}, the components $G_{rr} - \alpha T_{rr}=0$, $G_{r\varphi} - \alpha T_{r\varphi}=0$ and $G_{ru} - \alpha T_{ru}=0$ are the main equations. The scalar equation is the standard equation. $G_{u\varphi} - \alpha T_{u\varphi}=0$ and $G_{uu} - \alpha T_{uu}=0$ are the supplementary equations. Finally $G_{\varphi\varphi} - \alpha T_{\varphi\varphi}=0$ is the trivial equation.

\subsection{Solution space}
Once the scalar field is given as initial data in series expansion
\be
\phi(u,r,\varphi)=\sum_{a=1}^\infty\frac{\phi_a(u,\varphi)}{r^a},
\ee
the unknown functions $\beta,U,V$ can be solved explicitly. In asymptotic form, they are
\begin{multline}
\beta=\frac{3 \alpha \phi_1 \p_u \phi_1}{4r^3} + \frac{\alpha}{2r^4} \bigg[ 2M \phi_1^2 + 4 (\p_\varphi \phi_1)^2 - 2 \phi_1 \p_\varphi^2 \phi_1\\
 + 5 \phi_1^2 \p_u \phi_1 + 6 \phi_2 \p_u\phi_1 + 2 \phi_1 \p_u \phi_2 \bigg] + \cO(r^{-5}),
\end{multline}
\begin{multline}
U=\frac{N(u,\varphi)}{r^2}\\
- \frac{\alpha}{6r^4}\bigg[20 \p_u\phi_1 \p_\varphi \phi_1 + \phi_1 \left(3 \p_\varphi M \p_u \phi_1 - \p_u N \p_u \phi_1 - 4 \p_u\p_\varphi \phi_1\right)\bigg] + \cO(r^{-5}),
\end{multline}
\begin{multline}
V=-r M(u,\varphi) - \frac1r\left[ N^2 - 2 \alpha \p_u \phi_1 (2\p_u\phi_1 - \phi_1 \p_u M)\right]\\
 + \frac{\alpha}{3r^2} \bigg[4(1-M) \phi_1^2 \p_u M - 8 \phi_2 \p_u \phi_1 \p_u M - 4 \phi_1 \p_u \phi_2 \p_u M\\
 + \p_\varphi M \p_\varphi \phi_1 \p_u \phi_1 - 2 \p_\varphi \phi_1 \p_u N \p_u \phi_1 - 4 \p_\varphi^2 \phi_1 \p_u \phi_1 + 24 \p_u \phi_1 \p_u \phi_2\\
 + 16 \p_\varphi \phi_1 \p_u\p_\varphi \phi_1  + \phi_1 \bigg(16 M \p_u \phi_1 - 3 \p_\varphi M \p_u \phi_1 + 8 (\p_u \phi_1)^2 + 6 \p_u \phi_1 \p_u\p_\varphi N\\
  + \p_\varphi M \p_u \p_\varphi \phi_1 - 2 \p_u N \p_u\p_\varphi \phi_1 - 4 \p_u\p_\varphi^2 \phi_1 \bigg)\bigg] + \cO(r^{-3}),
\end{multline}
where $N(u,\varphi)$ and $M(u,\varphi)$ are integration constants. Compared to the pure Einstein case \cite{Barnich:2010eb}, the $\alpha$ corrections are at least two orders after the integration constants. The solution space is no longer in a closed form.

The time evolution of every order of the scalar field is controlled from the standard equation. This means that there is no news function from the scalar field. We list the first two orders of the standard equation
\begin{align}
&2(\p_u \phi_1)^2 + \phi_1 (\p_u M + 4 \p_u^2 \phi_1 )=0 ,\\
&4\p_u (\phi_1 \p_u \phi_2) + 8 \p_\varphi \phi_1 \p_u\p_\varphi \phi_1 + 12 \phi_2 \p_u^2 \phi_1 + 2 \phi_1^2 \p_u^2 \phi_1 -4 \p_\varphi^2 \phi_1 \p_u \phi_1 \nn\\
&\hspace{1cm} + \phi_1 \left(\p_\varphi^2 M + 8 M \p_u \phi_1 + 10 (\p_u \phi_1)^2 - 2 \p_u\p_\varphi N\right) - 2 \p_\varphi M \p_\varphi \phi_1 \nn\\
&\hspace{2cm} + 4 \p_\varphi \phi_1 \p_u N  + \frac52 \phi_1^2 \p_u M + 3 \phi_2 \p_u M=0 .
\end{align}
The constraints from the supplementary equations are
\begin{align}
&\p_u M=0,\\
&\p_u N =\frac12 \p_\varphi M,
\end{align}
which are the same as pure Einstein case. This is well expected as the $\alpha$ corrections are in the higher orders. In the end, there is no propagating degree of freedom at all in this theory in three dimensions. The whole effect of the higher dimensional Gauss-Bonnet terms is kind of a deformation of Einstein gravity.

\section{Asymptotic structure of Einstein-Gauss-Bonnet theory in four dimensions}
\label{sec:4d}

We now turn to more realistic four dimensions. The action is given by \eqref{lagrangian} with $p=4$. The derivation of the equations of motion is quite similar to the three dimensional case with the additional contribution from the Gauss-Bonnet term which is detailed in Appendix \ref{appendix}. The Einstein equation is obtained as
\be
G_{\mu\nu} - \alpha T_{\mu\nu}=0,
\ee
where the modifications to $T_{\mu\nu}$ \eqref{t} from the Gauss-Bonnet term is
\begin{multline}
-4 R_{\mu\rho\nu\sigma}\n^\rho \n^\sigma \phi + 4 G_{\mu\nu} \n^2 \phi - 4 R_\mu^\rho \n_\nu \n_\rho \phi - 4 R_\nu^\rho \n_\mu \n_\rho \phi\\
 + 4 g_{\mu\nu} R^{\rho\sigma} \n_\rho \n_\sigma \phi + 2 R \n_\mu\n_\nu \phi ,
\end{multline}
and the scalar equation is
\begin{multline}
G^{\mu\nu} \n_\mu \n_\nu \phi + R^{\mu\nu} \n_\mu \phi \n_\nu \phi  + \n^2\phi (\n \phi)^2  - (\n^2\phi)^2 + 2 \n_\rho \n_\sigma \phi \n^\sigma\phi \n^\rho \phi \\
 +  \n_\rho \n_\sigma \phi \n^\sigma \n^\rho \phi - \frac18 \left(R^{\mu\nu\rho\sigma}R_{\mu\nu\rho\sigma} - 4 R^{\mu\nu} R_{\mu\nu } + R^2\right)=0.
\end{multline}

\subsection{Bondi gauge}

In four dimensions, we choose the Bondi gauge fixing ansatz \cite{Bondi:1962px}
\begin{multline}\label{metric}
ds^2=\left[\frac{V}{r}e^{2\beta} + U^2r^2e^{2\gamma}\right]du^2 - 2e^{2\beta}dudr\\
 - 2U r^2e^{2\gamma}dud\theta + r^2\left[e^{2\gamma}d\theta^2 + e^{-2\gamma}\sin^2\theta d\phi^2\right],
\end{multline}
in $(u,r,\theta,\varphi)$ coordinates. The metric ansatz involves four functions $(V,U,\beta,\gamma)$ of $(u,r,\theta)$ that are to be determined by the equations of motion. These functions and the scalar field are $\varphi$-independent and hence the metric has manifest global Killing direction $\partial_\varphi$. This is the ``axisymmetric isolated system'' introduced in \cite{Bondi:1962px}.\footnote{In the present work, our main purpose is to demonstrate the effects of the Gauss-Bonnet term in the asymptotic analysis. For simplicity, we adopt the axisymmetric condition. However we do not expect any principle difficulties in the study of the general four dimensional asymptotic flatness solutions by choosing Sachs' gauge fixing ansatz \cite{Sachs:1962wk}.}
Following closely \cite{Bondi:1962px}, the falloff conditions for the functions $(\beta,\gamma,U,V)$ in the metric for asymptotic flatness are given by
\be\label{boundary1}
\beta=\cO(r^{-1}),\;\;\;\;\gamma=\cO(r^{-1}),\;\;\;\;U=\cO(r^{-2}),\;\;\;\;V=-r + \cO(1).
\ee
Considering the metric of Minkowski vacuum
\be
ds^2=-du^2-2dudr+r^2(d\theta^2 + \sin^2\theta d\phi^2),
\ee
we have two branches of the scalar solution
\be
\phi=0,\quad \text{or} \quad \phi=\log \frac{r}{r_0}.
\ee
The first gives the true vacuum with the maximal spacetime symmetry preserved; the second
solution is nearly Minkowski since the scalar does not preserve the full symmetry. Both are valid solutions, with one {\it not} encompassing the other. Analogous emergence of logarithmic dependence for the scalar also occurs in the anti-de Sitter (AdS) vacuum in some critical Einstein-Horndeski gravity, where the scalar breaks the full conformal symmetry of the AdS to the subgroup of the Poincare together with the scaling invariance \cite{Li:2018rgn}. However, ours is the first example in Minkowski vacuum. The necessary falloff condition of the scalar field consistent with the metric falloffs is either
\be\label{boundary2}
\phi=\cO(r^{-1}),\quad \text{or}\quad \phi=\log \frac{r}{r_0} + \cO(r^{-1}).
\ee

Similar to the three dimensional case, the equations of motions are organized as follows: $G_{rr} - \alpha T_{rr}=0$, $G_{r\theta} - \alpha T_{r\theta}=0$ and $G_{\theta\theta}g^{\theta\theta} + G_{\varphi\varphi}g^{\varphi\varphi} - \alpha T_{\theta\theta} g^{\theta\theta} - \alpha T_{\varphi\varphi} g^{\varphi\varphi}=0$ are the main equations. The scalar equation and $G_{\theta\theta} - \alpha T_{\theta\theta}=0$ are the standard equations. $G_{u\theta} - \alpha T_{u\theta}=0$ and $G_{uu} - \alpha T_{uu}=0$ are supplementary. $G_{ru} - \alpha T_{ru}=0$ is trivial. $G_{r\varphi} - \alpha T_{r\varphi}=0$, $G_{\theta\varphi} - \alpha T_{\theta\varphi}=0$ and $G_{u\varphi} - \alpha T_{u\varphi}=0$ are trivial because the system is $\varphi$-independent. 

\subsection{Solution space with the ordinary vacuum}

Supposing that $\gamma$ and $\phi$ are given in series expansion as initial data\footnote{To avoid logarithm terms in the metric, we turn off the order $\cO(r^{-2})$ in $\gamma$ following \cite{Bondi:1962px,Sachs:1962wk}.}
\begin{align}
\gamma&=\frac{c(u,\theta)}{r} + \sum^{\infty}_{a=3}\frac{\gamma_a(u,\theta)}{r^a},\\
\phi&=\sum^{\infty}_{a=1}\frac{\phi_a(u,\theta)}{r^a}.
\end{align}
The unknown function $\beta,U,V$ are solved out in asymptotic form as\footnote{Note that there is a shift in the integration constant $N(u,\theta)$ compared to the result in \cite{Lu:2019jus}.}
\begin{flalign}
&\beta=-\frac{c^2}{4r^2} + \frac{4 \alpha \phi_1 \p_u \phi_1}{3r^3} + \cO(r^{-4}),&
\end{flalign}
\begin{multline}
U=-\frac{2\cot\theta c + \p_\theta c}{r^2} + \frac{N(u,\varphi)}{r^3}\\
+ \frac{1}{2r^4}\bigg[5 \cot\theta c^3 - 3c N + 6 \cot\theta \gamma_3 + \frac52 c^2 \p_\theta c + 3 \p_\theta c\\
 + \alpha\left( 16 \cot\theta \phi_1 \p_u c - \frac{20}{3} \p_\theta \phi_1 \p_u \phi_1 + 8 \phi_1 \p_u\p_\theta c + \frac43 \phi_1 \p_u\p_\theta \phi_1\right) \bigg] + \cO(r^{-5}),
\end{multline}
\begin{multline}
V=-r + M(u,\theta) + \frac{1}{2r}\bigg[ \cot\theta N - \frac12 c^2 (5+11\cos2\theta)\csc^2\theta - 5 (\p_\theta c)^2 + \p_\theta N\\
 - c(19\cot\theta \p_\theta c + 3 \p_\theta^2 c) + 8 \alpha (\p_u\phi_1)^2 \bigg] + \cO(r^{-2}),
\end{multline}
where $N(u,\theta)$ and $M(u,\theta)$ are integration constants. Clearly, the coupling $\alpha$ emerges just one order after the integration constants. They are from non-minimal coupled scalar rather than the four dimensional Gauss-Bonnet term.

The standard equations control the time evolution of the initial data $\gamma$ and $\phi$. In particular, the time evolution of every order of the scalar field has been constrained. That means there is no news function associated to the scalar field. Hence the scalar field does not have propagating degree of freedom similar to the three dimensional case. We list the first two orders of the scalar equation
\begin{align}
&(\p_u\phi_1)^2 + \phi_1 \p_u^2 \phi_1=0,\\
&2 \phi_1 \p_u^2 \phi_2 + 6 \p_u \phi_1 \p_u \phi_2 - \p_\theta^2 \phi_1 \p_u \phi_1 - \cot\theta \p_\theta \phi_1 \p_u \phi_1 \nn\\
&\hspace{1cm} + 4 \p_\theta \phi_1 \p_u\p_\theta \phi_1 + \phi_1^2 \p_u^2 \phi_1 + 6 \phi_2 \p_u^2 \phi_1 + 6 \phi_1 \p_u \phi_1 + 4 \phi_1 (\p_u\phi_1)^2\nn\\
&\hspace{1cm} + \phi_1\left[\p_u\p_\theta^2 c + 3 \cot\theta \p_u\p_\theta c - 2 \p_u c - 2 (\p_u c)^2 - \p_u M \right]=0.
\end{align}
The first order of the standard equation from Einstein equation is
\begin{multline}
\p_u \gamma_3=\frac18 \bigg[3 (\p_\theta c)^2 + c (5\cot\theta \p_\theta c + 3 \p_\theta^2 c) -2 c^2 \csc^2\theta (3 + \cos2\theta)\\
+ 2 c M + \cot\theta N - \p_\theta N - 16 \alpha \phi_1 \p_u^2 c \bigg].
\end{multline}
In the Newman-Penrose variables, $\gamma_3$ is related to $\Psi_0^0$ or $\xbar\Psi_0^0$ \cite{Conde:2016rom}. Since its time evolution involves $\alpha$, the effect of the higher dimensional Gauss-Bonnet term arises starting from the first radiating source, i.e. quadrupole, in the multipole expansion \cite{Janis:1965tx}. This can be seen more precisely on the linearized level from the logarithm vacuum case that we will demonstrate in the next subsection.

The supplementary equations yield
\be
\p_u N = \frac13\left[ 7 \p_\theta c \p_u c + c(16\cot\theta \p_u c + 3 \p_u\p_\theta c) - \p_\theta M \right].
\ee
\be
\p_u m=-2(\p_u c)^2,\quad m \equiv M - \frac{1}{\sin\theta} \p_\theta ( 2 \cos\theta  c + \sin\theta \p_\theta c).
\ee
The latter is the mass-loss formula in this theory. It is the same as pure Einstein case \cite{Bondi:1962px} which is well expected as the corrections from the Gauss-Bonnet term are in the higher orders.

\subsection{Solution space with the logarithm vacuum}

One intriguing feature of the theory is that the scalar admits a logarithmic dependence in the Minkowski vacuum, such that the full Lorentz group breaks down for any matter coupled to the scalar.  We would like to analyse its solution space here. Supposing that $\gamma$ and $\phi$ are given in series expansion as initial data
\begin{align}
\gamma&=\frac{c(u,\theta)}{r} + \sum^{\infty}_{a=3}\frac{\gamma_a(u,\theta)}{r^a},\\
\phi&=\log \frac{r}{r_0} + \sum^{\infty}_{a=1}\frac{\phi_a(u,\theta)}{r^a}.
\end{align}
We can solve out the unknown function $\beta,U,V$ in asymptotic form as
\begin{multline}
\beta=-\frac{c^2}{4r^2} + \frac{1}{4r^4}\bigg[-3 c \gamma_3 + \alpha \bigg(4c\cot\theta (\p_\theta c + \p_\theta \phi_1 ) + c^2 ( \csc^2\theta + 3 \cot^2\theta) - \phi_1^2\\
 - 2 \phi_2 + (\p_\theta c)^2 + 2 \p_\theta c\p_\theta \phi_1 + (\p_\theta \phi_1 )^2 + 2 \alpha \p_u \phi_1 - 8 \alpha (\p_u \phi_1 )^3\bigg)\bigg] + \cO(r^{-5}),
\end{multline}
\begin{multline}
U=-\frac{2\cot\theta c + \p_\theta c}{r^2} + \frac{N(u,\varphi)}{r^3}\\
+ \frac{1}{2r^4}\bigg\{5 \cot\theta c^3 - 3c N + 6 \cot\theta \gamma_3 + \frac52 c^2 \p_\theta c + 3 \p_\theta c + \alpha \bigg[ 14\p_\theta c \p_u c \p_u \phi_1\\
 - 4 \p_\theta c \p_u c - 4 \p_\theta \phi_1 \p_u c - 4 \p_\theta c \p_u \phi_1 - 2 \p_\theta M \p_u \phi_1  - 6 \p_u N \p_u \phi_1\\
 - 2 c \left( 4\cot\theta \p_u c - 16 \cot\theta \p_u c \p_u \phi_1 + 4 \cot\theta \p_u \phi_1 - 3 \p_u \p_\theta c \p_u \phi_1 \right)\bigg] \bigg\} + \cO(r^{-5}),
\end{multline}
\begin{multline}
V=-r + M(u,\theta) + \frac{1}{2r}\bigg[ \cot\theta N - \frac12 c^2 (5+11\cos2\theta)\csc^2\theta - 5 (\p_\theta c)^2 + \p_\theta N\\
 - c(19\cot\theta \p_\theta c + 3 \p_\theta^2 c) - 2 \alpha + 8 \alpha (\p_u\phi_1)^2 \bigg] + \cO(r^{-2}).
\end{multline}
The coupling $\alpha$ emerges again one order after the integration constants. At this order, it is from non-minimally coupled scalar. The $\alpha^2$ terms in $\beta$ is indicating the nonlinear scalar-gravity coupling.

The time evolution of every order of the scalar field is also constrained. There is no news associated to the scalar field. The first two orders of the scalar equation are
\begin{align}
&\p_u \phi_1 + (\p_u \phi_1)^2 - (\p_u c)^2 - \frac12=0,\\
&4\phi_2 \p_u^2 \phi_1 - 4 \p_u \phi_2 - 8 \p_u \phi_1 \p_u \phi_2 - 3 M - 2 \phi_1 + 3 \cot\theta \p_\theta c + \cot\theta \p_\theta \phi_1 \nn\\
&\hspace{0.5cm} + \p_\theta^2 c + \p_\theta^2 \phi_1 - 2 \cot\theta \p_\theta c \p_u c + 2 \cot\theta \p_\theta \phi_1 \p_u c - 2 \p_\theta^2 c \p_u c - 2 \p_\theta^2 \phi_1 \p_u c \nn\\
&\hspace{0.5cm} - 4 \phi_1 (\p_u c)^2 - 6 \phi_1 \p_u \phi_1 - 12 \cot\theta \p_\theta c \p_u \phi_1 - 4\cot\theta \p_\theta \phi_1 \p_u \phi_1 - 4 \p_\theta^2 c \p_u \phi_1\nn\\
&\hspace{0.5cm} - 4 \p_\theta^2 \phi_1 \p_u \phi_1 + 8 \phi_1 (\p_u\phi_1)^2 + 4 \p_\theta c \p_u\p_\theta \phi_1 + 4 \p_\theta \phi_1 \p_u \p_\theta \phi_1 - 2c + 8 c \p_u \phi_1\nn\\
&\hspace{0.5cm} +c \p_u c (8\csc^2\theta - 4 \p_u \phi_1 ) + 8 \cot\theta c \p_u \p_\theta \phi_1 - 2 c^2 \p_u^2 \phi_1 + 2 \phi_1^2 \p_u^2 \phi_1 =0.
\end{align}
The first order of the standard equation from Einstein equation is
\begin{multline}
\p_u \gamma_3=\frac18 \bigg[3 (\p_\theta c)^2 + c (5\cot\theta \p_\theta c + 3 \p_\theta^2 c) -2 c^2 \csc^2\theta (3 + \cos2\theta)\\
+ 2 c M + \cot\theta N - \p_\theta N - 8 \alpha \p_u c + 16 \alpha \p_u \phi_1 \p_u c \bigg].
\end{multline}
The constraints from the supplementary equations are
\be
\p_u N = \frac13\left[ 7 \p_\theta c \p_u c + c(16\cot\theta \p_u c + 3 \p_u\p_\theta c) - \p_\theta M \right].
\ee
\be
\p_u m=-2(\p_u c)^2,\quad m \equiv M - \frac{1}{\sin\theta} \p_\theta ( 2 \cos\theta  c + \sin\theta \p_\theta c).
\ee
The mass-loss formula is the same as pure Einstein case \cite{Bondi:1962px}.

To reveal the $\alpha$ correction in radiating source, we linearize the theory for which we drop all the quadratic terms in the solutions. Then the evolution equations are reduced to
\begin{align}
&\p_u M = \frac{1}{\sin\theta} \p_\theta\left[\frac{1}{\sin\theta} \p_\theta \left(\sin^2\theta \p_u c\right)\right],\\
&\p_u N = -\frac13 \p_\theta M,\\
&\p_u \gamma_3=- \frac18 \sin\theta \p_\theta \frac{N}{\sin\theta} -  \alpha \p_u c.
\end{align}
The $\alpha$ correction now is only from the scalar background $\log \frac{r}{r_0}$ term. The multipole expansion is encoded in the expansion of $\gamma$ \cite{Janis:1965tx}. The quadrupole\footnote{To compare with eq.(2.46) of \cite{Janis:1965tx}, one has to adapt everything into Newman-Penrose formalism. We will not do that in the present work as our motivation is to reveal the $\alpha$ correction only.} in eq.(2.46) of \cite{Janis:1965tx} corresponds to $\gamma_3=a_2(u) \sin^2\theta$ where the subscript 2 denotes the second order of the second associated Legendre function. The function $c$ can be solved from above evolution equations. The solution is $c=c_2(u)\sin^2\theta$ where $c_2(u)$ satisfying
\be
c_2-\alpha \p_u^2 c_2=\p_u^2 a_2.
\ee
Suppose that $a_2$ are periodic functions e.g. $a_2=A \sin u + B \cos u$. Then the response of $c_2$ will have a $\alpha$ correction $c_2=\frac{\p_u^2 a_2}{1+\alpha}$. By setting $\alpha=0$, we just recover the Einstein gravity result $c=\p_u^2 a_2 \sin^2\theta$. For the same type of gravitational source, the new theory \eqref{lagrangian} is indeed distinguishable from Einstein gravity. Since $c$ function has a direct connection to the Weyl tensors \cite{Conde:2016rom}, we can expect a direct experimental test of the $\alpha$ corrections.


\section{Conclusion and discussion}
\label{sec:end}

In this paper, the asymptotic structures of three and four dimensional EGB gravity have been studied in Bondi-Sachs framework. It is shown from the solution space that there is no scalar propagator in both dimensions. The $\alpha$ corrections are discussed in details from both gravitational solutions and radiating sources perspectives.

There are still several open questions in the theory \eqref{lagrangian} that should be addressed in the future. There is no scalar propagator in the theory, but there are differential couplings between gravity and scalar field. The absence of the scalar propagator is likely to be consistent with observations, it is of interest to know how to construct gravity-scalar vertex without a scalar propagator \cite{Bonifacio:2020vbk}. A second interesting point is from the holography. In three dimensions, asymptotically flat gravitational theory has a holographic dual description \cite{Barnich:2010eb,Bagchi:2012xr}. It would be very meaningful to explore the dual theory of the three dimensional EGB gravity. Another question worths commenting is from the recent proposal of a triangle equivalence \cite{Strominger:2017zoo}. Since the change of $c$ function has $\alpha$ corrections for the same type of gravitational source, the gravitational memory receives the $\alpha$ correction \cite{Frauendiener}. In the context of the triangle relation, it is a very interesting question that if the soft graviton theorem and the asymptotic symmetry have $\alpha$ corrections as well.

\section*{Acknowledgements}
\label{sec:acknowledgements}


The authors thank Yue-Zhou Li and Xiaoning Wu for useful discussions. This work is supported in part by the NSFC (National Natural Science Foundation of China) under Grant No. 11935009. H.L. is also supported in part by NSFC Grant No. 11875200. P.M. is also supported in part by NSFC Grant No. 11905156.

\appendix

\section{Useful relations}
\label{appendix}

We list some useful relations that may help the readers who are less familiar with the variational principle involving Gauss-Bonnet term.

\noindent The Bianchi identity:
\be
\n_\mu {R_{\nu\sigma\rho}}^\nu + \n_\nu {R_{\sigma\mu\rho}}^\nu + \n_\sigma {R_{\mu\nu\rho}}^\nu =0.
\ee
Commutator of $\n$:
\be
(\n_\mu\n_\nu - \n_\nu\n_\mu) S^{\rho\sigma} = {R^{\rho}}_{\tau\mu\nu} S^{\tau\sigma} + {R^{\sigma}}_{\tau\mu\nu} S^{\rho\tau}.
\ee
Variations of some relevant quantities:
\begin{align}
&\delta \sqrt{-g}=-\frac12\sqrt{-g}g_{\mu\nu}\delta g^{\mu\nu},\\
&\delta \Gamma^\sigma_{\mu\nu} = -\frac12 \n^\sigma \delta g_{\mu\nu} - \frac12g_{\mu\tau}\n_\nu \delta g^{\sigma\tau} - \frac12 g_{\nu\tau} \n_\mu \delta g^{\sigma\tau},\\
&g^{\mu\nu}\delta \Gamma^\sigma_{\mu\nu} = \frac12 g_{\mu\nu} \n^\sigma \delta g^{\mu\nu} - \n_\mu \delta g^{\sigma\mu},\\
&\delta {R^\sigma}_{\mu\rho\nu}=\n_\rho \delta \Gamma^\sigma_{\mu\nu} - \n_\nu \delta \Gamma^\sigma_{\mu\rho},\\
&\delta R_{\mu\nu}=\frac12 \left(g_{\sigma\rho} \n_\mu \n_\nu \delta g^{\sigma\rho} - g_{\sigma\nu} \n_\rho\n_\mu \delta g^{\rho\sigma} - g_{\sigma\mu} \n_\rho \n_\nu \delta g^{\rho\sigma} - \n^2 \delta g_{\mu\nu} \right),\\
&\delta R=R_{\mu\nu} \delta g^{\mu\nu} + \n_\mu \left(g_{\sigma\rho} \n^\mu \delta g^{\sigma\rho} - \n_\nu \delta g^{\mu\nu}\right),\\
&\delta G^{\mu\nu}=\frac12 \left(g_{\sigma\rho} \n^\mu \n^\nu \delta g^{\sigma\rho} - \n_\sigma \n^\mu \delta g^{\sigma\nu} - \n_\sigma \n^\nu \delta g^{\sigma\mu} + \n^2 \delta g^{\mu\nu} \right) + R^\mu_\sigma \delta g^{\nu \sigma}\nn\\
&\quad\,\, + R^\nu_\sigma \delta g^{\mu \sigma} - \frac12 R \delta g^{\mu\nu} - \frac12 g^{\mu\nu} R_{\sigma\rho} \delta g^{\sigma\rho} - \frac12  g^{\mu\nu}g_{\sigma\rho} \n^2 \delta g^{\sigma\rho} + \frac12 g^{\mu\nu} \n_\rho \n_\sigma \delta g^{\rho\sigma},\\
&\delta R^2 = 2 R R_{\rho\sigma} \delta g^{\rho\sigma} + 2R \left(g_{\sigma\rho} \n^2 \delta g^{\sigma\rho} - \n_\mu \n_\nu \delta g^{\mu\nu}\right),\\
&\delta (R^{\sigma\mu\rho\nu} R_{\sigma\mu\rho\nu})=4R_{\sigma\mu\rho\nu}\n^\nu\n^\mu\delta g^{\rho\sigma} + 2 R_{\sigma\mu\rho\nu} {{R^\sigma}_\tau}^{\rho\nu} \delta g^{\mu\tau},\\
&\delta (R^{\mu\nu} R_{\mu\nu}) = R_{\rho\sigma} \n^2 \delta g^{\rho\sigma} -  R_{\mu\rho} \n_\sigma \n^\mu \delta g^{\rho\sigma}\nn\\
&\quad\quad\quad + g_{\rho\sigma} R^{\mu\nu} \n_\mu \n_\nu \delta g^{\rho\sigma} - R^\mu_\sigma \n_\mu \n_\rho \delta g^{\rho\sigma} + R_{\mu\nu} R^\nu_\sigma \delta g^{\mu\sigma} + R_{\sigma\mu\rho\nu} R^{\mu\nu} \delta g^{\sigma\rho},\\
&g^{\mu\nu} \delta (\n_\mu \n_\nu \phi) =g^{\mu\nu} \n_\mu \n_\nu \delta \phi -  \frac12 g_{\mu\nu} \n_\sigma \phi \n^\sigma \delta g^{\mu\nu} + \n_\sigma \phi \n_\mu \delta g^{\sigma\mu},
\end{align}
\begin{multline}\label{gb}
\delta \cG=2 R_{\sigma\mu\tau\nu} {R_\rho}^{\mu\tau\nu} \delta g^{\sigma\rho} +  2 R R_{\rho\sigma} \delta g^{\rho\sigma} - 4 R_{\rho\nu} R^\nu_\sigma \delta g^{\rho\sigma} - 4 R_{\sigma\mu\rho\nu} R^{\mu\nu} \delta g^{\sigma\rho}\\
4 {R_{\sigma\mu\rho}}^\nu \n_\nu\n^\mu\delta g^{\rho\sigma} + 4 {R_{\nu\sigma\rho}}^\nu \n_\mu \n^\mu \delta g^{\rho\sigma} + 4 {R_{\mu\nu\rho}}^\nu \n_\sigma \n^\mu \delta g^{\rho\sigma}\\
- 4 g_{\rho\sigma} G^{\mu\nu} \n_\mu \n_\nu \delta g^{\rho\sigma} + 4 G^\mu_\rho \n_\mu \n_\sigma \delta g^{\rho\sigma}.
\end{multline}
The first line of \eqref{gb} equals to $\frac12 g_{\sigma\rho}\cG \delta g^{\sigma\rho}$ in four dimensions. So they will not contribute to the equations of motion. When doing integration by part, the second line and the third line vanish automatically for pure Gauss-Bonnet term. However it will contribute when scalar field is coupled to Gauss-Bonnet term e.g. $\phi\cG$. The second line and the third line can be reorganized in the following way
\begin{multline}
4 {R_{\sigma\mu\rho\nu}} \n^\nu\n^\mu\delta g^{\rho\sigma} + 4 {R_{\nu\sigma\rho}}^\nu \n_\mu \n^\mu \delta g^{\rho\sigma} + 4 R^\mu_\rho \n_\sigma \n_\mu \delta g^{\rho\sigma}\\
- 4 g_{\rho\sigma} G^{\mu\nu} \n_\mu \n_\nu \delta g^{\rho\sigma} + 4 R^\mu_\rho \n_\mu \n_\sigma \delta g^{\rho\sigma} - 2 R \n_\sigma\n_\rho \delta g^{\rho\sigma}.
\end{multline}
The point of such reorganization is that the indexes of the two covariant derivative in every term are symmetric. When integrating by part for pure Gauss-Bonnet term, both covariant derivative give zero identically.

\providecommand{\href}[2]{#2}\begingroup\raggedright\endgroup

\end{document}